# Dissipation in Nanocrystalline-Diamond Nanomechanical Resonators

A.B. Hutchinson, P.A. Truitt, and K.C. Schwab[a]

*Laboratory for Physical Sciences, University of Maryland, College Park, Maryland 20740*

L. Sekaric,[b] J. M. Parpia, and H. G. Craighead

*Cornell Center for Materials Research, Cornell University, Ithaca, New York 14853*

J. E. Butler

*Naval Research Laboratory, Washington, DC 20375*

We have measured the dissipation and frequency of nanocrystalline-diamond nanomechanical resonators with resonant frequencies between 13.7 MHz and 157.3 MHz, over a temperature range of 1.4-274 K. Using both magnetomotive network analysis and a novel time-domain ring-down technique, we have found the dissipation in this material to have a temperature dependence roughly following $T^{0.2}$, with $Q^{-1} \approx 10^{-4}$ at low temperatures. The frequency dependence of a large dissipation feature at ~35-55 K is consistent with thermal activation over a 0.02 eV barrier with an attempt frequency of 10 GHz.

[a] Electronic mail: schwab@lps.umd.edu
[b] Current address: IBM T.J. Watson Research Center, Yorktown Heights, NY 10598

Diamond as a mechanical material offers the largest known ratio of $\sqrt{E/\rho}$, where $E$ is Young's modulus and $\rho$ is density, resulting in the highest possible resonant frequency $f_0$ for a given device geometry. This property, combined with the highest thermal conductivity and chemical resistance, makes diamond an attractive material for nanoelectromechanical systems (NEMS). Recently, Sekaric et al.[1] have demonstrated nanomechanical resonators in nanocrystalline diamond at room temperature, with frequencies of 6-640 MHz and quality factors $Q$ of 2400-3500.

In addition to resonant frequency, dissipation $Q^{-1}$ is a critical factor in NEMS performance. The ultimate performance in applications such as microscopic force detection,[2] signal processing,[3] and fundamental quantum mechanical experiments[4,5] is directly related to the dissipation. To illuminate the possible dissipation mechanisms, we have studied the temperature dependence of $Q^{-1}$ for four diamond resonators with frequencies of 13.7 MHz, 55.1 MHz, 110.1 MHz, and 157.3 MHz, between 1.4 K and 274 K.

Doubly-clamped beams were fabricated as described in Ref. 1, with a width $w$ of 180 nm, a thickness of roughly 40 nm, and lengths $l$ = 2.5, 3, 4 and 8 μm. A 35 nm layer of gold was deposited on top, to allow magnetomotive actuation and detection[6] in fields up to $B$ = 9 T. The sheet resistance of this film was 1.6 Ω/□ between 1.4 K and ~20 K, rising linearly above 20 K to a value of 2.7 Ω/□ at 250 K. The samples were placed in high vacuum (< $10^{-5}$ Torr) on a 4-He cryostat with a base temperature of 1.4 K. The temperature was monitored using a carbon-glass thermometer and regulated using a resistive heater, both mounted on the copper sample stage. Given the large temperature dependence of the resonant frequency (Fig. 3), temperature stability of 1 mK was necessary for these measurements.

We measured the quality factor and resonant frequency using two methods that rely on magnetomotive effects. First, we used conventional network analyzer techniques. Although this approach is very convenient for identifying nanomechanical resonances, extraction of accurate quality factors relies on a detailed model of the embedding and stray impedances. In order to verify our fitting procedures, we have developed an alternate time-domain technique which is not sensitive to these impedances. We apply a 10,000-cycle pulse train at the resonant frequency, which excites the mechanical oscillation of the beam. The voltage developed across the beam is amplified, sampled and stored by a high-speed storage scope, which is triggered at the end of the pulse train. The left-hand inset of Fig. 1 shows the captured free decay, which is then processed using a fast Fourier transform (FFT). Figure 1 shows a comparison of the FFT power spectrum of the free decay data with network analysis data. Agreement between the two techniques for the extracted quality factor is better than 5%, giving us confidence in our network analysis fits.

Figure 2 shows the central result of this Letter: the dissipation $Q^{-1}$ of the diamond resonators as a function of temperature. In order to understand this data we have considered a variety of possible dissipation mechanisms: magnetomotive damping,[7] thermoelastic dissipation,[8,9] mechanical dissipation in the gold electrode,[10] clamping losses,[11] and finally dissipation in the diamond itself. We will address each mechanism below.

Magnetomotive damping, which results from electronic dissipation in resistive elements due to induced eddy currents, is a substantial effect for these[7] and other[12] devices. This mechanism is additive to the "intrinsic" dissipation $Q_0^{-1}$, and is a function of the electromechanical resistance of the beam, $R_{em}=(Bl)^2 Q_0 / 2\pi f_0 m$, and the impedance of the external measuring circuit, $Z_{ext}$:

$$Q^{-1} = Q_0^{-1}\left[1 + R_{em}\Re(Z_{ext})/|Z_{ext}|^2\right], \qquad (1)$$

where $B$ is magnetic field and $m$ is resonator mass.[7] Here $Z_{ext}$ includes the series combination of the resistance of the gold electrode on the resonator (~20-60 Ω), the impedance of the 1.5 m semi-rigid cable, and the input impedance of the preamplifier (50 Ω or 3 kΩ). The right-hand inset in Fig. 1 shows both the measured dissipation versus magnetic field and a fit using the above parameters, with the cable impedance approximated by a series inductance of 400 nH and a parallel capacitance of 110 pF. Comparable data shown by Mohanty *et al.*,[13] which was attributed to the presence of charged defects or electronic donor impurities, can be adequately fit with similar parameters. For each resonator studied here, we have measured $Q^{-1}$ versus $B$ and determined the largest field such that the magnetomotive damping contributes less than 5% of the total dissipation.

We have estimated the contribution of thermoelastic loss and find this to be negligible for all of the resonators presented here, over the entire temperature range studied. This results from the small width of the beam (180 nm) and the high thermal conductivity of diamond, which combine to make the thermal time constant of the beam several orders of magnitude shorter than the mechanical period.[8]

Another contribution to the dissipation is the internal losses of the gold film. We find that for a doubly-clamped beam composed of two materials,[14]

$$Q^{-1} = \frac{1}{1+\beta}\left(Q_{Dia}^{-1} + \beta Q_{Au}^{-1}\right), \quad (2)$$

where $\beta = t_{Au}E_{Au}/t_{Dia}E_{Dia} \approx 0.08$ in our devices. $Q_i$, $t_i$, and $E_i$ are the quality factor, beam thickness, and Young's modulus for diamond and gold. Using Eq. (2) and the measured dissipation in thin films of gold[10], we estimate that the contribution of the gold to the measured dissipation is less than 5% of the observed dissipation.

The dissipation of the shortest (157.3 MHz, $l = 2.5$ μm) beam displays an additive offset compared to the other three beams. This offset could be due to clamping loss, which was modeled for cantilevers by Jimbo and Itao.[11] A similar calculation for doubly-clamped beams gives the result $Q_{Clamp}^{-1} = 10.4(w/l)^3$. For the 2.5 μm beam, $Q_{Clamp}^{-1} = 3.9 \times 10^{-3}$, over 30 times greater than in the 8 μm beam. However, the prediction is clearly too large compared to the observed data, which indicates that the simple two-dimensional model used is not sufficient to capture the behavior of the beams. Given that this loss mechanism is predicted to be increasingly significant for high-frequency resonators, a more complete model would be very useful.

Except for the additive offset noted above in the 157.3 MHz resonator, the dissipation shown in Fig. 2 is believed to be from the diamond itself. The temperature dependence, although complex, is similar to that observed in GaAs,[13] Si,[8,15] SiO$_2$,[14] and metallic films such as Au and Al.[10] This apparent universality in both magnitude and temperature dependence provides further evidence that the dissipation arises from a broad spectrum of tunneling states, similar to those found in amorphous solids.[10] The overall trend in the low-temperature data roughly follows a power law with exponent ~0.2, which has yet to be explained.

A prominent feature seen in this data, as with other materials,[8,13,15] is a Debye peak at ~35-55 K. The inset to Fig. 2 shows an Arrhenius plot of the position of this peak, which displays a linear dependence. This indicates a thermally activated process with energy $\Delta E \approx 0.02$ eV and an attempt frequency $\Gamma_0 \approx 10$ GHz. A more dramatic peak is observed at T = 220 K in the 13.7 MHz resonator. It is difficult to speculate on the nature of this peak, as it was not observed in the other beams over the temperature range accessible by our apparatus.

Figure 3 shows the temperature dependence of resonator frequency, which is similar in magnitude and sign to shifts observed in Si and GaAs.[13,15] However, the increasing frequency at higher temperatures (> 160 K) in the 13.7 MHz and the 110.1 MHz resonators is new behavior. (The 55.1 MHz resonator was destroyed before we were able to perform measurements at higher temperatures.) Although we suspect this effect is a result of relative thermal expansion, calculating the size of thermal strain is complicated by the complex structure of the samples: three thin films (Au, diamond, SiO$_2$) with unknown stresses, all atop a Si substrate. A simple model incorporating only the length change and tension produced in the resonator by the thermal expansion of the Si substrate and diamond film predicts a minimum in $\delta f / f$ around 140-160 K, with the appropriate order of magnitude. However, the deviation at low temperatures indicates that factors not yet understood also play a significant role. Interestingly, the magnitude of the frequency shift does not depend monotonically on resonator length, which may be a result of mesoscopic variations of the mechanical properties from sample to sample. This might be expected, since the diamond grain size (~5-40 nm) can be a significant fraction of the resonator width (180 nm).

In summary, we have measured the dissipation of nanocrystalline diamond nanomechanical resonators between 13.7 MHz and 157.3 MHz. We find that the magnitude and temperature dependence of $Q^{-1}$ are very similar to those observed in other materials, suggesting a similar mechanism for dissipation. The power law $Q^{-1} \sim T^{0.2}$ observed at low temperatures here and with other materials begs further study and understanding. To help answer these questions, measurements of this and other materials, including single-crystal diamond, at ultra-low temperatures (10 mK - 1 K) would be very illuminating. In addition, by combining the time-domain technique demonstrated here with very-high-sensitivity readout amplifiers,[16] it may also be possible to observe discrete dissipation events. This would offer further insight into the mesoscopic behavior of these nanomechanical devices.

The authors thank Olivier Buu, Art Vandelay, Elinor Irish and Hidehiro Yoshida for fruitful discussions. This work was funded by the National Security Agency, DARPA/MTO, and the National Science Foundation.

**FIGURE CAPTIONS**

FIG. 1. The solid line shows the resonance of the 13.7 MHz beam, measured using network analysis. The filled circles show the results of an FFT of the ring-down transient; the raw time-domain data, averaged 500 times, is shown in the left-hand inset. The right-hand inset shows the measured dissipation (open circles) as a function of magnetic field, along with the calculated (solid line) magnetomotive dissipation.

FIG. 2. Dissipation as a function of temperature for the four resonators. Data for the 13.7 MHz resonator, which was taken at $B = 9$ T, has been adjusted to remove the effect of magnetomotive damping, using Eq. (1). (The other resonators were measured at low enough field that magnetomotive damping was negligible.) The arrows indicate the position of a Debye peak; the inset shows an Arrhenius plot of $\ln(f)$ versus $1/T$ for this peak.

FIG. 3. The shift in resonant frequency $\delta f / f_0$ as a function of temperature for the four resonators, plotted on a logarithmic scale. The solid and dotted lines show the predictions of a model based on the thermal expansion of Si and diamond.

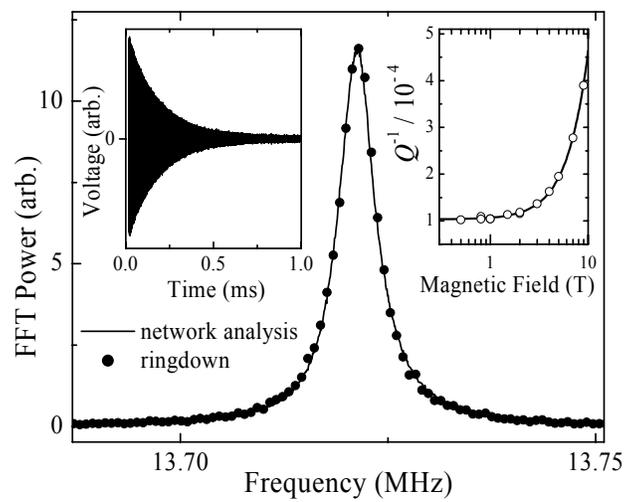

FIG. 1. Hutchinson *et al.*, APL

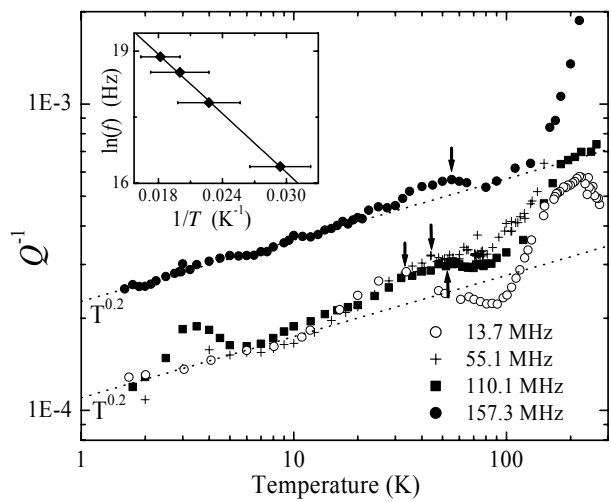

FIG. 2. Hutchinson *et al.*, APL

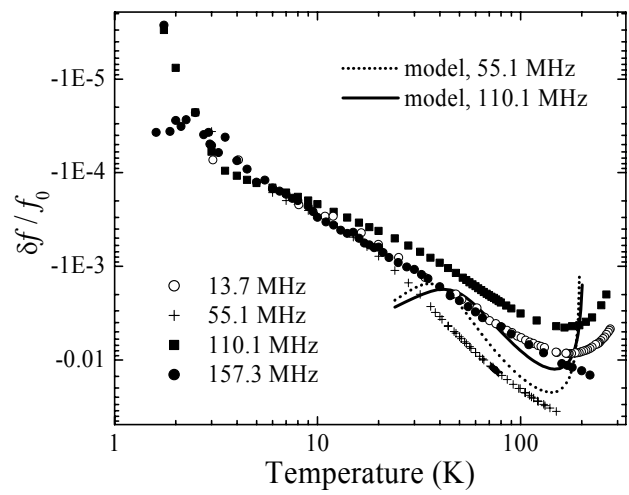

FIG. 3. *Hutchinson et al.*, APL